\documentclass[twocolumn,prl,aps,superscriptaddress,showpacs]{revtex4}

\usepackage{mathptmx}
\usepackage{subfigure}
\usepackage{dcolumn}
\usepackage{amsmath,amssymb}
\usepackage{bm}
\usepackage{color}
\usepackage{overpic}
\usepackage{latexsym}
\usepackage{epstopdf}
\usepackage{color}
\usepackage[english]{babel}
\usepackage{latexsym}

\usepackage{psfrag,graphicx} 
\usepackage{epsf} 
\usepackage{subfigure} 
\usepackage{amsmath} 
\usepackage{amssymb} 
\usepackage{amsfonts}
\usepackage{bm}
\usepackage{natbib}
\usepackage{epstopdf}
\usepackage{appendix}

\definecolor{mygrey}{gray}{0.35}
\definecolor{myblue}{rgb}{0.,0.,1}
\definecolor{myzard}{cmyk}{0,0,0.05,0}
\definecolor{mywhite}{rgb}{1,1,1}
\definecolor{myred}{rgb}{1,0.,0.3}

\usepackage[colorlinks=true,citecolor=blue,linkcolor=blue]{hyperref}

\def\be{\begin{equation}}
\def\ee{\end{equation}}
\def\ba{\begin{align}}
\def\enda{\end{align}}
\def\bi{\begin{itemize}}
\def\ei{\end{itemize}}

 \def\ee{\mathord{\rm e}}
 
 \def\ii{\mathord{\rm i}}

\def\half{\textstyle\frac{1}{2}}

 \def\ee{\mathord{\rm e}}
 
 \def\ii{\mathord{\rm i}}

\def\half{\textstyle\frac{1}{2}}

\renewcommand{\ii}{{\rm i}}
\renewcommand{\ee}{{\rm e}}

\def\beq{\begin{equation}}
\def\beq{\begin{equation}}
\def\eeq{\end{equation}}

 \newcommand{\ket}[1]{|#1\rangle}
 \newcommand{\bra}[1]{\langle #1|}

\begin{document}

\title[Short Title]{Localization of phonons in ion traps with controlled quantum disorder}

\author{A. Bermudez}
\affiliation{Departamento de F\'{i}sica Te\'orica I,
Universidad Complutense, 28040 Madrid, Spain }
\author{M.A. Martin-Delgado}
\affiliation{Departamento de F\'{i}sica Te\'orica I,
Universidad Complutense, 28040 Madrid, Spain }
\author{D. Porras}
\affiliation{Departamento de F\'{i}sica Te\'orica I,
Universidad Complutense, 28040 Madrid, Spain }

\begin{abstract}
We show that the vibrations of a chain of trapped ions offer an interesting route to explore the physics of disordered quantum systems.
By preparing the internal state of the ions in a quantum superposition, we show how the  local  vibrational energy becomes a stochastic variable, being its statistical properties inherited from the underlying quantum parallelism of the internal state. We describe a  minimally-perturbing measurement of the resonance fluorescence, which allows us to study effects like Anderson localization without the need of  ground-state cooling or individual addressing,  and thus paves the way towards high-temperature ion experiments.

\end{abstract}
\maketitle

Disorder in quantum systems leads to a variety of fascinating phenomena. 
In particular,
the disorder-induced localization of particles  has been at the focus of intense theoretical and experimental research since its discovery by P.W. Anderson \cite{anderson}.
Experiments have been performed in a variety of setups  involving electronic transport \cite{al_review}, 
propagation of light \cite{al_light} 
or sound waves \cite{al_ultrasound}, and disordered Bose-Einstein condensates 
\cite{al_bec,review_palencia}.
So far, any experiment has limitations in the degree of control of particle interactions or the statistical properties of disorder. Thus, it remains elusive to find realizations of most challenging situations, such as Anderson localization in the presence of correlated disorder or controlled interactions. In this work we show that the vibrations of an ion crystal can be used to study the physics of Anderson localization. 
The ability to control the interactions and measure the state of trapped ions 
\cite{wineland_review},  
makes this setup an ideal quantum simulator of many-body systems (see e.g.~\cite{porras_spin,porras_phonon_hubbard,retzker_zig_zag,singer_polariton}), as recently shown in experiments \cite{simulator,monroe_spin}. 
Our proposal relies on the description of the radial vibrations of an ion chain in terms of weakly coupled harmonic oscillators \cite{porras_phonon_hubbard}. 
These vibrations are coupled by means of lasers to the internal states of the ions, represented by a set of two-level systems, or effective spins. We employ a state-dependent Stark-shift that  shifts the local ion trapping potential, to couple spins and phonons. This composite system can be described effectively as two different species interacting in a chain.
Since the spin species is the slowest, it behaves like a frozen background for phonons, and makes the local phonon energy  a true stochastic variable whose statistical properties are determined by the spin quantum state.

In particular, we show that: {\it i)} By preparing the ion internal state in a quantum superposition corresponding to a separable state, disorder leads to the localization of  phonons. This effect can be directly measured by studying the propagation of phonons along the chain. 
{\it ii)} Fundamental properties, like the phonon localization length, may be detected even at high temperatures, and without the need to resolve individual ions. This is achieved by measuring the resonance fluorescence. {\it iii)} Our  proposal can be extended to implement a variety of interesting situations, such as binary-correlated disorder, or localization in the presence of anharmonicities.

{\it General scheme.-} 
We consider a chain of $N$ ions with mass $m$ and charge $e$ trapped by harmonic potentials. 
The ions are coupled by the Coulomb interaction, and thus, their displacements around the equilibrium positions are described by collective vibrational modes. 
In particular, we focus on the vibrations in the radial -- transverse to the chain --  direction. 
Each ion also has two internal levels ($\ket{0}, \ket{1}$) with energy difference $\omega_0$. 
Thus, the system Hamiltonian is (we set $\hbar = 1$)
\beq
H_0=\omega_0\sum_{j=1}^N\sigma_j^z+\sum_{n=1}^{N}\Omega_n a_{n}^{\dagger}a_{n},\hspace{2ex} \Omega_{n}^2 =\omega_{\text{t}}^2 \left( 1+\beta\mathcal{V}_n \right),
\eeq
where $\sigma_j^z = \ket{0_j}\bra{0_j}-\ket{1_j}\bra{1_j}$, $a^{\dagger}_{n}$ ($a_{n}$) create (annihilate)  radial phonons in the vibrational mode $n$, and
$\Omega_n$ are the  radial-mode energies determined by the trapping frequency $\omega_{\rm t}$, and the harmonic correction ${\cal V}_n$. The later is obtained from the diagonalization of $V_{jk} =|j-k|^{-3}(1-\delta_{jk}) - \sum_{l\neq j}|j-l|^{-3}\delta_{jk}$, such that $\mathcal{V}_n = \sum_{jk}{\cal M}_{j n}V_{jk}{\cal M}_{k n}$, where ${\cal M}_{j n}$ are the phonon wavefunctions. 
Finally, the ratio of the Coulomb repulsion to the trapping energy  $\beta = e^2/m\omega_{\text{t}}^2a^3$ fulfills $\beta \ll 1$, such that the ions correspond to weakly coupled oscillators \cite{porras_phonon_hubbard}. We work under the simplification of equally spaced ions, although our results can be easily extended to inhomogeneous chains. 

We use a laser in the radial direction with  frequency   $\omega_{\rm L}$ and wavevector $k_{\rm L}$, which is nearly resonant to a narrow transition between the internal states. Thus, the vibrational sidebands can be resolved, and the laser couples spins and phonons through the spatial dependence of the dipole coupling 
$H_{\rm L} = \frac{\Omega_{\rm L}}{2}\sigma_j^+ e^{\ii k_{\rm L} x_j - \ii \omega_{\rm L} t} + {\rm H.c.}$, where $\sigma_j^+ = \ket{1_j}\bra{0_j}$.
In the limit $\beta \ll 1$, $\omega_{\rm L}$ can be tuned near resonance with all the modes by choosing $\omega_{\text{L}} - \omega_0 = -\Omega_n + \delta_n$, and detunings $\delta_n \ll \Omega_n$.
Expressing the radial coordinate as 
$x_j = \sum_n {\cal M}_{j n} (a_n + a^\dagger_n)/ \sqrt{2 m \Omega_n}$, the coupling can be approximated by a multimode red-sideband Hamiltonian. Here, one considers   $\Omega_{\text{L}}\ll \Omega_n$ and
$\eta_n = k_{\text{L}}/\sqrt{2m\Omega_n}\ll1$, where the carrier and remaining sidebands can be neglected, and the Hamiltonian  in the interaction picture with respect to  $H_0$, reads
\beq
\label{sideband}
 H_{\text{int}}(t)=\sum_{jn}F_{jn}\sigma_j^+a_n\ee^{-\ii\delta_nt}+\text{H.c.},
\hspace{2ex} F_{jn} = \frac{\ii\Omega_{\text{L}}}{2}\eta_n\mathcal{M}_{jn} .
\eeq  
Under the condition $|F_{jn}| \ll \delta_n$, we get an effective Hamiltonian from  the corrections to second order in 
$|F_{jn}|/\delta_n$,
\beq
\label{adiab_ham}
H_{\text{eff}}=H_0+\sum_{jkn}\lambda_{jkn}\sigma_j^+\sigma_k^- +\sum_{jnm}\kappa_{jnm}\sigma_j^za_n^{\dagger}a_m,
\eeq
where $\lambda_{jkn} = 2F_{jn}F^*_{kn}/\delta_n$ is an effective coupling mediated by the vibrational modes, and 
$\kappa_{jnm} = -F_{jn}F^*_{jm}(\delta_n+\delta_m)/2\delta_n\delta_m$ stands for the spin-phonon coupling. 
Eq. (\ref{adiab_ham}) can be recast in terms of local modes, 
$a_j = \half\sum_n\mathcal{M}_{jn}(\Delta_{n}^+a_n+\Delta_{n}^-a_n^{\dagger})$, 
with $\Delta_{n}^{\pm}=(\frac{\omega_{\text{t}}}{\Omega_n})^{\frac{1}{2}}\pm({\frac{\Omega_n}{\omega_{\text{t}}}})^{\frac{1}{2}}$. In the limit $\beta \ll 1$,  
the Hamiltonian can be split into three different terms 
$H_{\text{eff}}=H_{\text{p}} + H_{\text{s}}+H_{\text{s-p}}$,
\beq
\label{local_hamiltonian}
\begin{split}
&H_{\text{p}} = \sum_j \omega_j a_j^{\dagger}a_j + \sum_{jk} t_{jk}a_j^{\dagger}a_k,\\
&H_{\text{s}} = \sum_j\omega_0\sigma_j^z+\sum_{jk}J_{jk}\sigma_j^+\sigma_k^-, 
\hspace{2ex} H_{\text{s-p}} = U \sum_j \sigma_j^z a_j^{\dagger} a_j.
\end{split}
\eeq
The last expression has a clear physical meaning:
$H_{\text{p}}$ is an effective tight-binding Hamiltonian describing the harmonic coupling between ions in terms of phonon hopping;
$H_{\text{s}}$ stands for a dipolar XY-model of spins subjected to a longitudinal field; and $H_{\text{s-p}}$ is the spin-phonon coupling induced by a state-dependent Stark-shift of the local ion energy \cite{schmidt.kaler}. For non-resolved sidebands, this term would correspond to a spin-dependent light force that does not involve the phonon modes~\cite{ivanov}. Typical values for the coupling constants in (\ref{local_hamiltonian}) depend on
$F=\Omega_{\text{L}}\eta_{\text{L}}/2$, 
$\eta_{\text{L}}=k_{\text{L}}/\sqrt{2m\omega_{\text{t}}}$,
and $\delta=\omega_{\text{L}}-\omega_0+\omega_{\text{t}}$ (see Table~\ref{tabla}).

\begin{table}[!hbp]
\caption{Spin, phonon, and spin-phonon coupling strengths.}
\begin{ruledtabular}
\begin{tabular}{l l l}
\hspace{1.7ex}$\omega_{j}=\omega_{\rm t}-\sum_{l\neq j} \frac{t}{|j-l|^3}$ & \hspace{-0.4ex}$\omega_j\sim\omega_{\rm t}$ & \hspace{-0.7ex}$\omega_{\rm t}\sim10^4\text{kHz}$\hspace{2ex} \\
\hspace{2ex}$t_{jk}=\frac{t}{|j-k|^3}$ & $t=\beta\omega_t$ & $t\hspace{0.2ex}\sim\hspace{0.8ex}10^2\text{kHz}$\hspace{2ex} \\
\hspace{2ex}$U_{j}=U$ & $U=-\frac{F^2}{\delta}$ & $U\sim10^{\hspace{1ex}}\text{kHz}$\hspace{2ex} \\
\hspace{2ex}$J_{jk}=\frac{J}{|j-k|^3}$ & $J=\frac{F^2}{\delta^2}\beta\omega_{\rm t}$ & $J\hspace{0.2ex}\sim\hspace{0.8ex}1\hspace{0.2ex}\text{kHz}$\hspace{2ex} 
\end{tabular}
\end{ruledtabular}
\label{tabla}
\end{table}
The Hamiltonian in Eq.\eqref{local_hamiltonian} corresponds to a chain with two species (phonons and spins), coupled by a density--density interaction. 
By choosing  $\omega_{\text{t}}\sim10$ MHz, $\delta\sim1\text{MHz}$, $\beta~\sim0.05$, 
and $F\sim0.1\delta$, we obtain the hierarchy of coupling strengths 
$t\sim 100\text{kHz}\gg U\sim 10\text{kHz}\gg J\sim1\text{kHz}$. 
Accordingly, the spin dynamics is much slower than the typical phonon timescales, such that spins may serve as an auxiliary system to induce a disordered background for phonons. Note that the separation in time scales is a consequence of the fact that we consider the limit $(F/\delta)^2 << 1$. In this limit, induced effective spin-spin interactions are suppressed with respect to the phonon couplings, $J = (F/\delta)^2 t$. A similar situation has also been studied for atom-mixtures in optical lattices~\cite{paredes_disorder}, whereas the effect of randomness in quantum networks was addressed in~\cite{torma}.

\begin{figure}[!hbp]
  \centering
  \begin{overpic}[width=6.50cm]{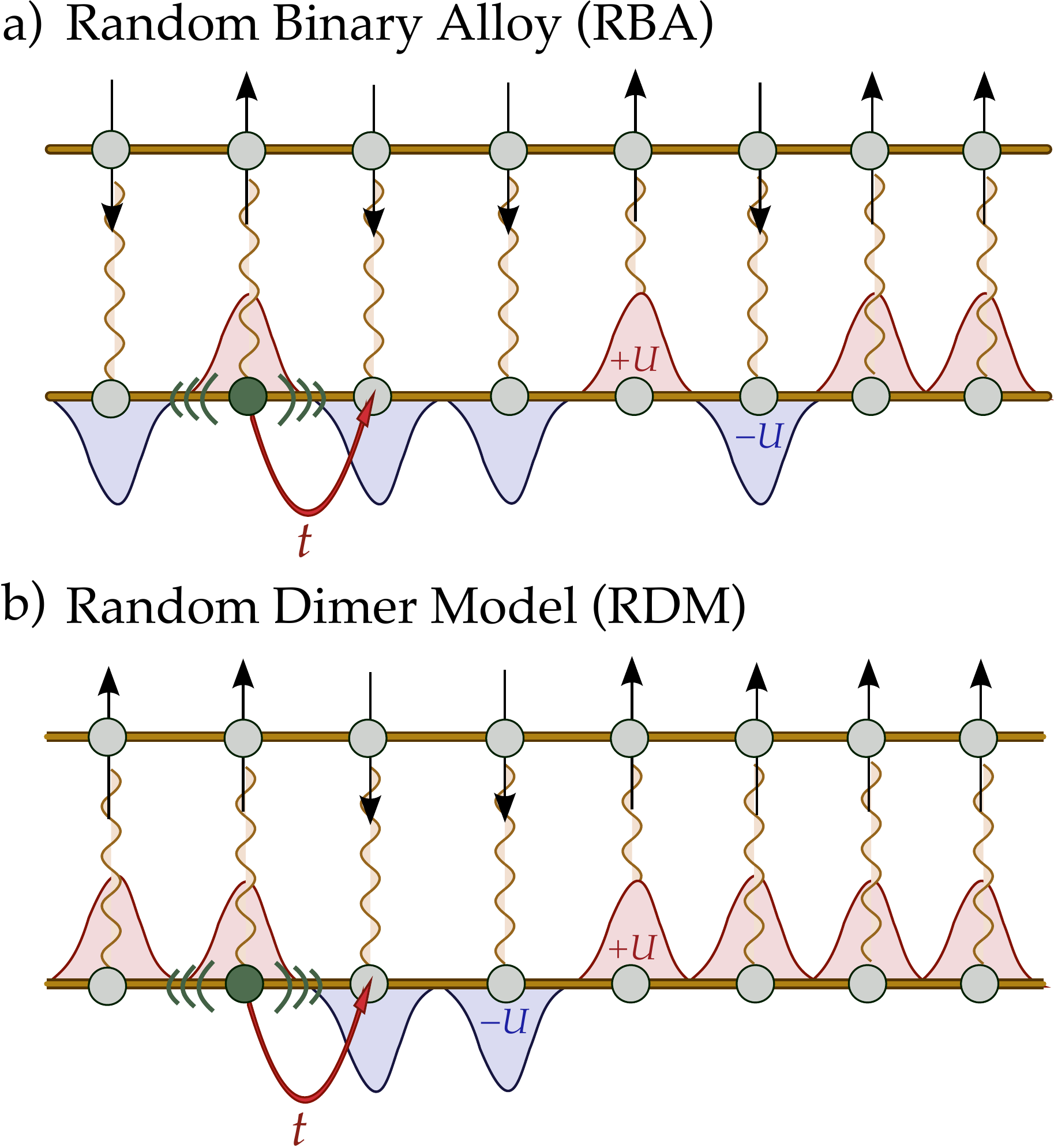} \end{overpic}
 \caption{Disordered ion chain as a virtual two-leg ladder composed of a frozen-spin ensemble locally coupled to a system of hopping  phonons. a) Uncorrelated RBA. b) Correlated  RDM.}
 \label{disordered_chain}
 \end{figure}

{\it Controlled disorder.-} 
We consider the initial state 
$\rho(0) = \rho_{\text{s}}^0 \otimes \rho_{\text{p}}^0$, 
where 
$\rho_{\text{s}}^0=\ket{\Psi_{\text{s}}}\bra{\Psi_{\text{s}}}$ 
is an arbitrary pure state with
$\ket{\Psi_{\text{s}}} =
\sum_{\{\boldsymbol{s}\}}c_{\{\boldsymbol{s}\}}\ket{\{\boldsymbol{s}\}}$.
The set $\{\boldsymbol{s}\}=\{s_1, \dots, s_N\}$, with $s_{j}\in\{0,1\}$,
defines a given spin configuration, and
$c_{\{\boldsymbol{s}\}}\in\mathbb{C}$ are the corresponding probability amplitudes. 
Due to the quenching of the spin dynamics, the phonon evolution can be approximated by
$\rho_{\text{p}}(t)
=\sum_{\{\boldsymbol{s}\}}|c_{\{\boldsymbol{s}\}}\hspace{-0.2ex}|^2\ee^{-\ii H_{\text{eff}}^{\{\boldsymbol{s}\}}t}\rho^0_{\text{p}}\ee^{\ii H_{\text{eff}}^{\{\boldsymbol{s}\}}t}$, 
where
\beq
\label{RBA_ham}
 H_{\text{eff}}^{\{\boldsymbol{s}\}}=H_{\text{p}}+H_{\text{s-p}}^{\{\boldsymbol{s}\}}=\sum_{jk}t_{jk}a_j^{\dagger}a_k+\sum_j\epsilon_{j}(\{\boldsymbol{s}\})a_j^{\dagger}a_j,
\eeq
and $\epsilon_{j}(\{\boldsymbol{s}\})=\omega_j+U(1+2s_j)$ is an on-site  energy with two possible values $\epsilon_{j} = \omega_{j}\pm U$. 
Due to the underlying quantum parallelism, the system explores simultaneously all  realizations of the on-site energy  with  probabilities  $p(\epsilon_{j}(\{\boldsymbol{s}\}))=|c_{\{\boldsymbol{s}\}}\hspace{-0.2ex}|^2$. We stress that the  binary randomness inherited from the frozen-spin background  can be engineered at will preparing different internal states $\ket{\Psi_{\text{s}}}$. In principle, arbitrary quantum states can be prepared by trapped-ion quantum logic
\cite{ion_gates}.
Therefore, we obtain a versatile phonon model with  diagonal disorder,  formally equivalent to the so-called random binary alloy (RBA)~\cite{random_binary_alloy}, but in stark contrast, we can externally  control the  disorder (see Fig.~\ref{disordered_chain}(a)). Note that the long-range  hopping  in Eq.~\eqref{local_hamiltonian} does not  change qualitatively the phase diagram of the model.

{\it Anderson localization in the RBA model.-} 
In  the absence of disorder, phonon wavefunctions are extended over the whole ion chain.  
Conversely, impurities lead to interference phenomena of back-scattered waves that might localize the phonon wavefunctions.  This effect, known as Anderson localization, occurs for any amount of disorder in 1D systems ~\cite{mott}.  In contrast to impurities that may consist of ions with a different mass~\cite{ivanov}, we consider here the spin ensemble as a source of disorder. As an illustrative example, we focus on the RBA model with uncorrelated disorder, and  choose a separable initial spin state  $\rho_s^0=\bigotimes_j{\ket{+_j}}\bra{+_j}$, with $\ket{+_j}=\left(\ket{0_j}+\ket{1_j}\right)/\sqrt{2}$. The latter can be prepared by means of a laser or microwave field addressed to the internal transition of each ion~\cite{wineland_review}. 
Note that such state leads to a flat probability distribution of the local phonon energies in Eq.~\eqref{RBA_ham}, which is characterized by $p(\epsilon_{j}(\{\boldsymbol{s}\}))=1/2^N$. 
In order to observe localization of phonons, we can either study the dynamics of  single phonon excitations, or the spectral properties of a thermal distribution of phonons.  Below, we describe these two possible routes. 

\begin{figure}[!hbp]
 \centering
 \begin{overpic}[width=7.0cm]{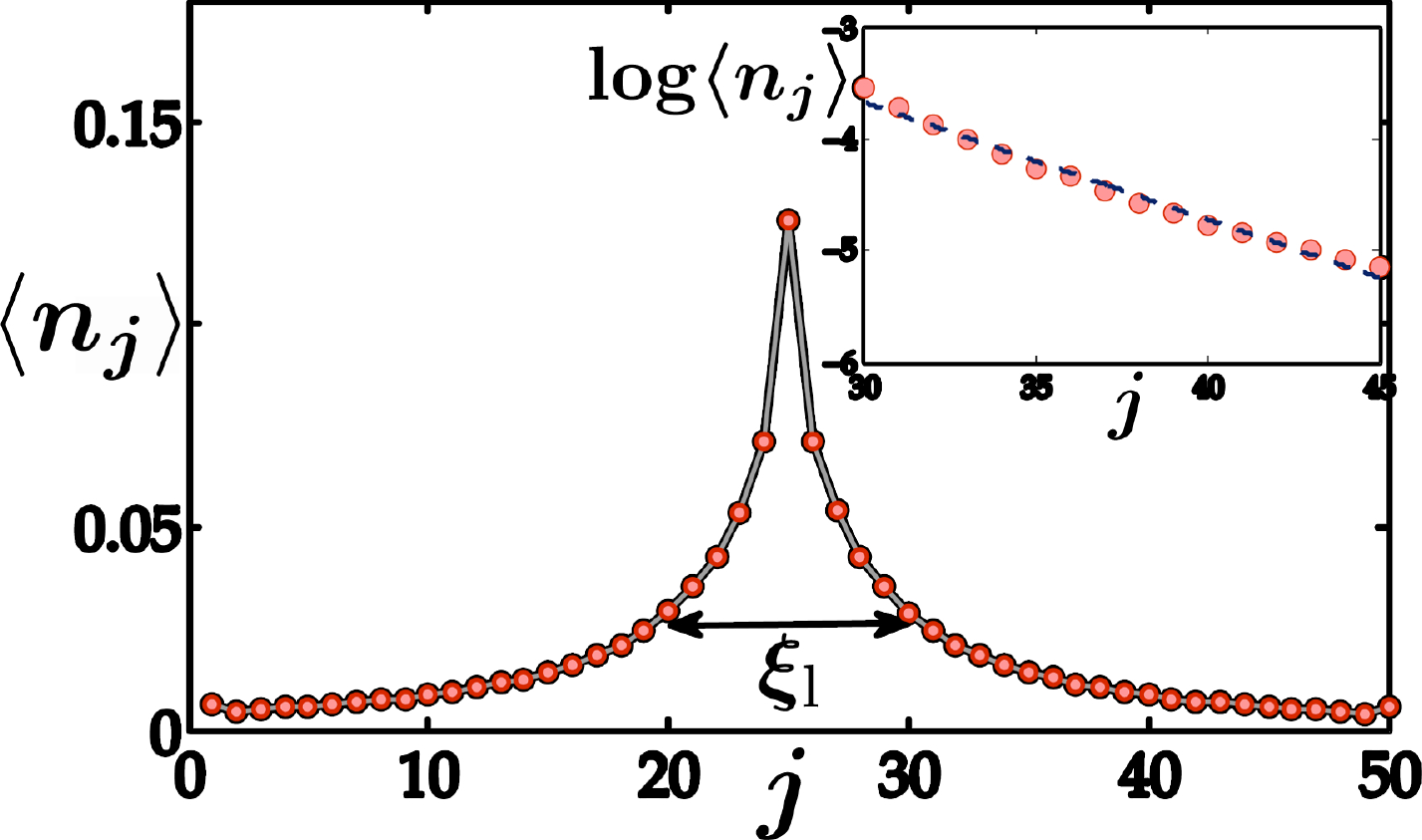}
 \end{overpic}
 \caption{Mean phonon number $\langle n_j(t_{\text{f}})\rangle$ at $t_{\text{f}}=10^3/\beta\omega_{\text{t}}$, for a disordered chain, $U=3t/4$, with $N=50$ ions. A localized phonon at the center of the chain  does not spread completely, but remains localized within the length-scale $\xi_{\text{l}}$. Inset: exponential tail of the distribution which gives us the   localization length, $\xi_{\text{l}}=A^{-1}\approx 10\ll N$.}
 \label{dyn}
 \end{figure}

{\it a) Direct measurement of localization.--} 
The required experimental techniques to create and measure the quantum state of motion have been demonstrated with single ions~\cite{meekhof}. In this case, we also need cooling close to the vibrational ground state and individual addressing of ions. To create a single phonon in a ion on the chain, we assume an experimental procedure which is faster that the timescale for phonon tunneling, $1/t$. This is in principle feasible, since sidebands can be resolved on time scales longer than $1/\omega_t$. Considering a time scale $t_p$ for the pulse sequence required to create a single phonon, and typical values $\omega_t = 10 $MHz, and $t = 100$ kHz, the required condition $1/\omega_t \ll t_p \ll 1/t$ may be fulfilled. Note that this is in agreement with time scales shown in~\cite{meekhof}.
The experiment should proceed as follows: 
{\it i)} Laser cooling close to the ground state, where the required  mean phonon number $\bar{n} \ll 1$ may be achieved by sideband cooling without individual resolution of the vibrational modes. 
{\it ii)} Initialization, where a single phonon localized at a given site $l$ is created by a red-sideband coupling (single-ion version of Eq. (\ref{sideband})) that drives the transition 
$\ket{1_l}\otimes\ket{\text{vac}}_{\text{p}} \to \ket{0_l}\otimes a_l^{\dagger}\ket{\text{vac}}_{\text{p}}$, where $\ket{\text{vac}}_{\text{p}}$ is the phonon vacuum. 
Afterwards, internal states should be prepared in the coherent superposition $\ket{+_j}$ 
{\it iii)} Switch on the disorder on-site phonon energies in (\ref{RBA_ham}), and
wait for single-phonon propagation along the chain until a given final time $t_\text{f}$. {\it iv)} Infer the phonon number by  measuring the excitation probability of internal states.
In the absence of disorder, the long-time dynamics at $t_{\text{f}} \gg \text{max}\{1/\Omega_n\}$ corresponds to a quantum random walk where the phonon diffuses over the entire chain. Conversely, disorder yields a drastically different behavior as shown in the numerical results of Fig.~\ref{dyn}, where phonons are confined within a localization length $\xi_{\text{l}}\approx10$ sites.

\begin{figure}[!hbp]
 \centering
  \begin{overpic}[width=7cm]{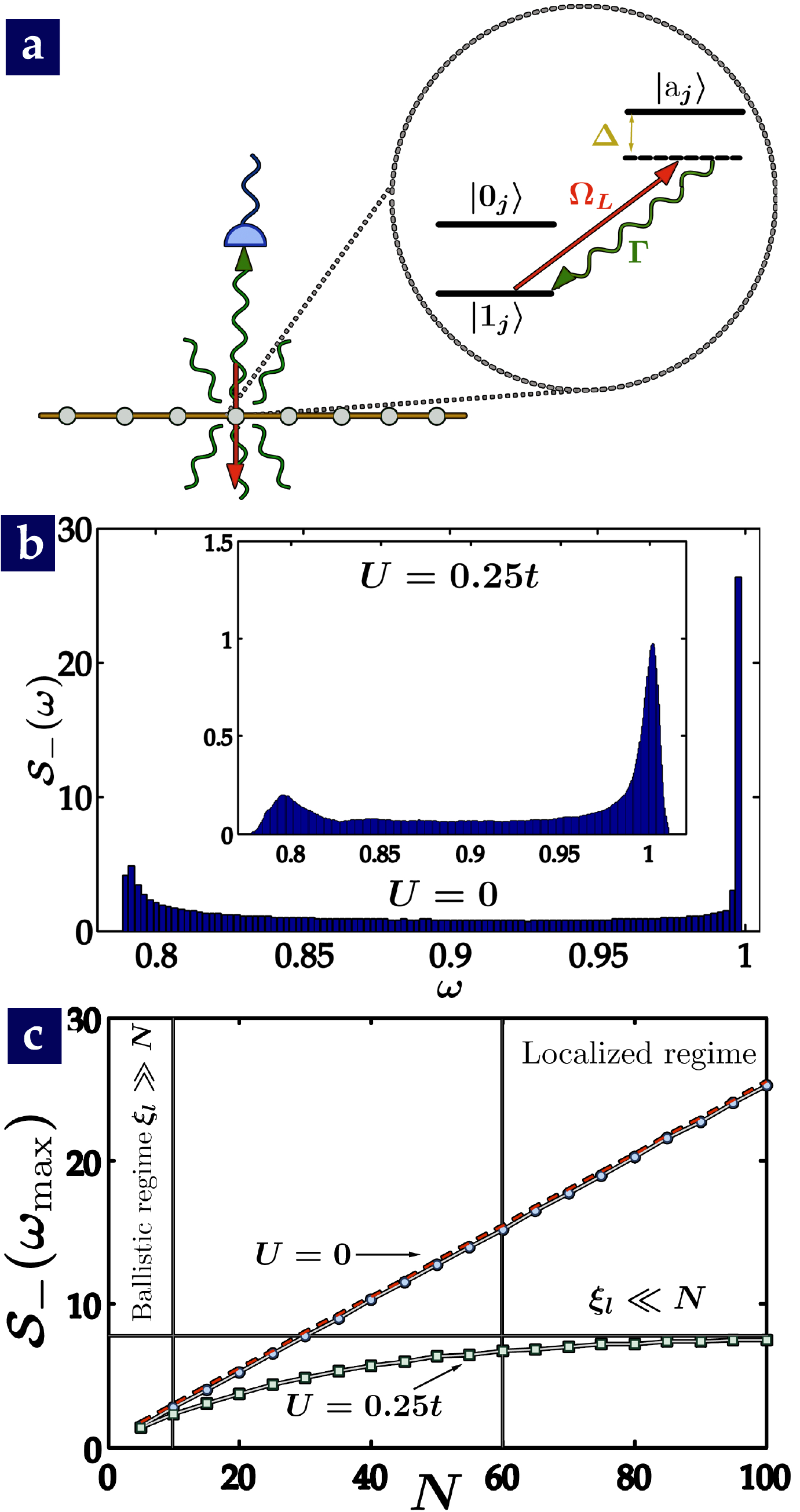}
 \end{overpic}
 \caption{(a) Scheme of the cycling transition coupling  the ion selectively $\ket{1_j}$ to  $\ket{\text{a}_j}$. Selection rules impose the decay back to $\ket{1_j}$. (b) Fluorescence $\mathcal{S}_-(\omega)$ (units of $\mathcal{S}_0^-$) as a function of $\omega$ (units of $\omega_t$) for a chain of $N=100$  and $\beta=0.05$. Main figure: ordered case $U=0$. Inset: disordered regime $U/t=0.25$. 
 (c) Scaling of the fluorescence for  $\beta=0.05$. 
 Blue circles: Ordered regime $U=0$, which agrees with $\mathcal{S}_-(\Omega_{\text{com}})\propto \frac{1}{4}(1+N)$ (red dashed line).
 Green squares: Disorder $U/t=0.25$, where  saturation shows phonon localization.}
 \label{scheme_fluo}
 \end{figure}

{\it b) Spectral measurement of localization.--}
{
In the following we show how to measure phonon spectral properties.
Our scheme perturbs minimally the spins, such that we remain in the quenched disorder limit implicit in Eq. (\ref{RBA_ham}). 
For this aim, we propose to measure the fluorescence from a cycling transition, and show how to  
get the phonon localization length out of it, without neither cooling to the ground state nor individual addressing.
To describe the measurement scheme, we extend the theory of resonance fluorescence from a single atom 
\cite{cirac_cooling_spectrum} to an ion chain. 
A laser with wavevector 
${\bf k}_\text{c}$ in the radial direction drives a cycling transition between levels $| 1_j \rangle$ and $| a_j \rangle$, 
with detuning $\Delta$ and Rabi frequency $\Omega_\textmd{c}$. 
Photons with momentum $-{\bf k}_\text{c}$ are detected (see Fig. \ref{scheme_fluo}a)).
We consider the limit $\Gamma \gg \Delta \gg \Omega_c/2$, 
where $\Gamma$ is the natural linewidth of the cycling transition. 
In this limit, the excited level $| a_j \rangle$ can be adiabatically eliminated
and we get the following coupling between the ions and the fluorescence photons,
\beq
\label{effective.cycling}
 \begin{split}
  &H_{\text{cyc}}(t)= 
\sum_{j, \textbf{q} \approx {\bf k}_\text{c}}
g_{\textbf{q}}O_j c_{\textbf{q}}^{\dagger}\ee^{\ii(\omega_{\textbf{q}}-\omega_\textmd{L})t}+\text{H.c.},\\
O_j &= \ii \frac{\Omega_{\rm c}}{\Gamma}  e^{i 2 k_{\textmd{L}} x_j} \ket{1_j}\bra{1_j} \\
    &= \ii \frac{\Omega_{\rm c}}{\Gamma}\left(1+2\ii\eta_\textmd{c} (a_j + a_j^{\dagger})\right)\ket{1_j}\bra{1_j} +  O(\eta_\textmd{c}^2),
  \end{split}
\eeq
where $c_{\textbf{q}}^{\dagger}$ ($c_{\textbf{q}}$) create (annihilate) photons with momentum $\textbf{q}$, $g_{\textbf{q}}$ is the dipole coupling of the cycling transition, and $\eta_\textmd{c} = k_\textmd{c} /\sqrt{2 m \omega_\textmd{t}}$.
Eq. (\ref{effective.cycling}) describes photon emission weighted by $\Omega_\textmd{c}/\Gamma$, the probability amplitude for occupation of 
$| a_j \rangle$. The fluorescence spectrum is given by $\mathcal{S}(\omega) = 
\lim_{T\to\infty} \frac{1}{2\pi T} 
\int_0^T \text{d}t \int_0^T \text{d}\tau\sum_{jl} \langle O_j^{\dagger}(t)O_l(t + \tau)\rangle\ee^{\ii(\omega - \omega_{\text{L}})\tau}$,
(see for example \cite{Scully_book}).
Near $\omega = \omega_{\rm L} \pm \omega_{\rm t}$ the fluorescence is given by a sum of contributions from the different vibrational modes.

To be specific, we compute the resonance fluorescence spectrum from a phonon distribution at fine temperature $T$ after Doppler cooling. In this limit, collective vibrational modes are not resolved, and thus the phonon density matrix describes a set of individual harmonic oscillators,
$\rho_{\text{p}} =
\mathcal{Z}^{-1}\text{exp}(-\frac{\omega_{\text{t}}}{T}\sum_ja_j^{\dagger}a_j)$. The fluorescence spectrum is then
\begin{equation}
\label{fluorescence}
\mathcal{S_{\pm}}(\omega)=\mathcal{S}_0^{\pm}\sum_{\{\boldsymbol{s}\}}|c_{\{\boldsymbol{s}\}}|^2\sum_{jln}s_js_l\mathcal{M}^{\{\boldsymbol s\}}_{jn}\mathcal{M}^{\{\boldsymbol s\}}_{ln}\delta(\omega-\omega_{\text{L}}\pm\Omega^{\{\boldsymbol s\}}_n),
\end{equation}
where 
$\mathcal{S}_0^+ = \mathcal{S}_0 \bar{n}$, 
$\mathcal{S}_0^-:= \mathcal{S}_0 (\bar{n}+1)$. 
Up to now, we have neglected the broadening of the spectral delta functions by the heating rates. This effect is discussed below. }
 
Let us show how
Eq. (\ref{fluorescence}) allows us to determine the phonon localization length. 
We focus on the center-of-mass (COM) mode ($n = 0$), which has the highest energy.
In the absence of disorder, the fluorescence COM peak intensity  
fulfills $\sum_{\{\boldsymbol s\}}|c_{\{\boldsymbol s\}}|^2\sum_{jl}s_js_l\mathcal{M}_{j0}\mathcal{M}_{l0}=\frac{1}{4}(N+1)$, because
$\mathcal{M}_{j 0} = 1/\sqrt{N}$.
Therefore the peak at the COM frequency scales linearly with the number of ions, $\mathcal{S}_-(\omega)\propto N\delta(\omega-\omega_L-\Omega_{\text{com}})$. 
Conversely, for any amount of disorder, wavefunctions become exponentially localized \cite{mott}. Since the wavefunction of phonons close to the COM frequency have now a spatial extent of the order of the localization length, $\xi_l$, we expect the COM fluorescence peak to saturate to 
$\mathcal{S}_-(\omega)\propto \xi_l \delta(\omega-\omega_L-\Omega_{\text{com}})$, when $N \gg \xi_l$.
Thus, by studying the scaling of the COM peak with the number of ions in the chain, the saturation of the intensity will reveal the Anderson localization length. 
We illustrate our method with a numerical calculation  in Figs.~\ref{scheme_fluo}(b) and~\ref{scheme_fluo}(c), 
where the blue-sideband fluorescence is represented with and without disorder. Localization is evident already with a moderate ion number  $N \approx 20$. 

{
We discuss now a few points on the measurement of resonance fluorescence. Note that the sideband contributions to the resonance fluorescence spectrum have a linewidth of the order of the cooling/heating rates, $\Gamma_{h/c}$, as shown, for example, in~\cite{cirac_cooling_spectrum}. The sideband spectrum can be obtained with an heterodyne detection scheme, as discussed in \cite{raab}. During the measurement, radial motion of the ion chain may be heated or cooled with a rate of the order of 
$\Gamma_{\text{h/c}} = \eta_\text{L}^2 \Omega_\text{c}^2  / \Gamma$. In a typical cycling transition we find $\Gamma \gg \omega_x$. By choosing $\Delta = \omega_x$, 
the experiment is in the Doppler cooling regime, such that in the steady state the mean phonon number has typical values $\bar{n} = 10$ \cite{wineland_review}. Thus, the resonance fluorescence can be measured with a steady phonon number within the Lamb-Dicke regime.
Besides that, Doppler cooling leads to the broadening of vibrational sidebands in 
Eq. (\ref{fluorescence}). However, this effect does not spoil the measurement, as long as $\Gamma_{\text{h/c}}$ is of the order of the separation between vibrational modes, in which case, broadening will not affect significantly the scaling of photoluminescence. Finally, note that decoherence of the internal states is allowed in our proposed experiment, since quantum coherence is not required for the effective spins to induce a disordered potential in Eq. (\ref{RBA_ham}). }

{\it Localization-delocalization transition.-} 
The localization paradigm is altered by the presence of strong interactions or 
correlated  disorder, which lead to  a variety of phases where a 
localization-delocalization transition holds. In the following, we show how to use our scheme to realize those situations.  

{\it a) Correlated disorder.-} 
In addition to the uncorrelated disorder discussed above, 
entanglement in $\ket{\Psi_{\text{s}}}$ leads to correlated disorder in the phonon Hamiltonian.
The simplest situation consists of local vibrational energies which take on two possible values $\epsilon_j = \omega_j\pm U$, with the constraint 
$\epsilon_j = \epsilon_{j+1}$ for $j$ odd (see Fig.~\ref{disordered_chain}(b)).
This is known as the random dimer model (RDM) \cite{random_dimer_model}.
To induce such correlations, one should use quantum gates to prepare Bell-pairs 
 $\ket{\Psi_{\text{s}}}=\bigotimes_{j=\text{odd}}\ket{\Phi^+_j}$, where $\ket{\Phi_j^+}=(\ket{0_j}\ket{0_{j+1}}+\ket{1_j}\ket{1_{j+1}})/\sqrt{2}$. 
Interestingly, this model presents a quantum phase transition between an insulating 
phase ($\nu:=t/U<1)$, and a phase with extended phonon wavefunctions ($\nu>1$), being  the 
critical point at $\nu_c=1$ for a nearest-neighbor model. 
Additionally, the disorder-averaged local density of states (LDOS) has a  different behavior in the two regimes~\cite{dos_disorder}. 
Thus, such phase transition can be detected in the phonon LDOS 
$\rho_j(\omega)=\langle\sum_n\mathcal{M}_{jn}^2\delta(\omega-\omega_L-\Omega_n)\rangle$, by means of the fluorescence spectrum when the  laser is  focused onto a single ion $\mathcal{S}_-(\omega)\propto\rho_j(\omega)$. 
We emphasize that  
ion traps  not only allows an ideal realization of the RDM, but suggests the 
possibility of engineering a wide range of disorder-correlations that  surpass  real materials.

{\it b) Bose-glass phase.-} 
Finally,  radial anharmonicities in the  trapping potentials can be enhanced by an additional standing-wave~\cite{porras_phonon_hubbard}. These non-linearities lead to repulsive (attractive) on-site interactions between the phonons whenever the ions sit at the maxima (minima) of the standing-wave laser. Combining these interactions with the effective Hamiltonian in Eq.~\eqref{RBA_ham}, one reaches the disordered strongly correlated model
\beq
\label{BG_ham}
 H_{\text{eff}}^{\{\boldsymbol{s}\}}=\sum_{jk}t_{jk}a_j^{\dagger}a_k+\sum_j\epsilon_{j}(\{\boldsymbol{s}\})a_j^{\dagger}a_j+\sum_jU(a_j^{\dagger})^2a_j^2,
\eeq
where the interaction stregth $U$ can be controlled by the laser parameters~\cite{porras_phonon_hubbard}. Accordingly, it is possible to study the fate of Anderson localization in the presence of disorder, where a yet to be observed insulating phase arises: the Bose glass~\cite{bose_glass,review_palencia}. In contrast to the Mott insulator that arises due to the suppression of density fluctuations in the strong repulsive regime, the Bose glass is a gapless and compressible phase, but still an insulator due to  disorder localization.  Trapped ions are to be considered as an attractive platform to realize and detect these different phases. In fact, they are well suited to distingish experimentally between Mott-insulating and  Bose-glass phases. Following~\cite{bose_glass_detection},  it is possible to infer the compressibility via measurements of the boson distribution under modifications of a confining potential. For trapped ions, the phonons are  harmonically confined  to the center of the Coulomb crystal, which  can be modified by tuning the trapping frequency or the ion number~\cite{porras_phonon_hubbard}. Under these variations, the phonon distribution can be measured using standard  techniques~\cite{wineland_review}, and thus the compressibility inferred. 

 {\it Conclusions.-}
 We have presented a scheme that uses ion crystals to measure phonon localization induced by disorder. We have shown that  measuring  the collective resonance fluorescence suffices to prove Anderson localization, which can be surprisingly observed even at high phonon temperatures.

{\it Acknowledgments.-} We acknowledge  support 
from  EU grant PICC (Physics with Ion Coulomb Crystals), QUITEMAD S2009-ESP-1594, FIS2009-10061, CAM-UCM/910758, FPU MEC grant, and RyC Y200200074. D. Porras thanks A.V. Malyshev, K. Singer and J. Eschner for discussions.


\end{document}